# Superfluidity in the Solar Interior: Implications for Solar Eruptions and Climate

**Oliver K. Manuel[1], Barry W. Ninham[2], and Stig E. Friberg[3]**

---

Efforts to understand unusual weather or abrupt changes in climate have been plagued by deficiencies of the standard solar model (SSM) [1]. While it assumes that our primary source of energy began as a homogeneous ball of hydrogen (H) with a steady, well-behaved H-fusion reactor at its core, observations instead reveal a very heterogeneous, dynamic Sun. As examples, the upward acceleration and departure of $H^+$ ions from the surface of the quiet Sun and abrupt climatic changes, including geomagnetic reversals and periodic magnetic storms that eject material from the solar surface are not explained by the SSM. The present magnetic fields are probably deep-seated remnants of very ancient origin. These could have been generated from two mechanisms. These are: a) Bose-Einstein condensation [2] of iron-rich, zero-spin material into a rotating, superfluid, superconductor surrounding the solar core and/or b) superfluidity and quantized vortices in nucleon-paired Fermions at the core [3].

---

[1]University of Missouri, Rolla, MO 65401 USA

[2] University of Florence, 50019 Florence, Italy and Australian National University, Canberra, Australia 0200

[3] Clarkson University, Potsdam, NY 13699 USA

# I. INTRODUCTION

Neutrons and protons, with spin 1/2, satisfy Fermi Dirac statistics. Matter comprised of fermions becomes more nearly perfect as density increases [4]. This observation led to the suggestion more than half a century ago that a collapsed supernova can undergo a transition from an ordinary star into a neutron star [5] and to predictions that a neutron star is stable only if its mass is $1/3 \, M_o < m < 3/4 \, M_o$ [6], where $M_o$ is one solar mass. By contrast the most abundant, most stable, nuclei that occur in astrophysical systems as a result of stellar evolution have zero or even spin and satisfy Bose Einstein statistics. And, like the charged fermi gas, a charged bose gas also becomes more nearly perfect with increasing density and temperature. Such a high density charged Bose fluid that occurs in astrophysical conditions can then under appropriate conditions undergo Bose-Einstein condensation. It becomes a superfluid superconductor [2]. The Meissner effect subsequently leads to expulsion of the magnetic field generated by collapse of the rotating, massive object. The field would be confined to the neighborhood of the superfluid surface.

Unlike for the Fermi fluid problem [3], these observations remarked on 40 years ago in [2] have been overlooked for astrophysical systems. Giant gaseous planets and the solar surface are mostly [1]H, a fermion, but the inner planets and the interior of the Sun consist mostly of bosons (abundant isotopes of Fe, Ni, O, Si, S, Mg and Ca) [7-11]. Hence, a reasonable conclusion is that the solar core consists of degenerate fermions in a neutron star [12,13] surrounded by a dense iron-rich core of a Bose Einstein superfluid, superconductor [2]. As a result, neutron-emission in the core may initiate a series of reactions that produce the Sun's luminosity, solar neutrinos, and the continuous upward flow of $H^+$ ions that maintains mass separation in the Sun and annually releases $3 \times 10^{43}$ $H^+$ ions from the surface in the solar wind [14,15]. We will show that deep-seated magnetic fields associated with superfluidity of nucleon-paired fermions in the solar core and/or Bose-Einstein condensation in material surrounding that core may explain the upward acceleration and departure of $H^+$ ions in the solar wind and abrupt climatic changes, including geomagnetic reversals and the periodic magnetic storms that mark the solar cycle by violently ejecting material from the solar surface.

## II.    THE EARTH-SUN CONNECTION

Life is fragile. Mankind lives in fear of calamity on the surface of a tiny, iron-rich planet that comprises about 0.0003% of the mass of the solar system. To calm these fears, public funds are channeled to the scientific community to explain the occurrence of natural events. The results are not always reassuring, e.g., witness the current debate over global warming. Climatic changes cause "*water shortages, crop damage, streamflow reduction, and depletion of groundwater and soil moisture*" [16]. "*Paleomagnetic investigations (augmented by geological, paleobiological, and geochronological studies) and magnetometer measurements of the ocean floor have established that the Earth's magnetic field reverses polarity frequently, but quite irregularly, with an average time between reversals of about 200,000 years*" [ref. 17, p. 456; 18].

The Sun comprises 99.9% of the mass of the entire solar system. The separation between the Earth and the Sun is <3% of the distance to the outermost planets. Not surprisingly, the Sun dominates most events on Earth, including our climate [16]. In fact unnatural processes driven by the solar energy and the solar-wind and solar-flare particles that wash over our planet sustain life, fossil fuels, and hydroelectric power. Solar magnetic fields are linked with solar activity [19], with the solar wind, with solar flares, and with many other parameters, e.g., the Van Allen radiation belts, the aurora borealis (northern lights), the aurora australis (southern lights), the North Atlantic Oscillations, sea level pressures, the stratospheric geopotential height, and the geomagnetic index [20]. It has been difficult to determine cause-and-effect between the Sun and the Earth because the Sun influences so many other parameters that affect our climate [20].

The most massive planet, Jupiter, comprises $\approx 0.1\%$ of the solar system. It is 5 times further from the Sun and about 300 times more massive than Earth. It contributes to the Sun's irregular oscillation about the solar system's center-of-mass, producing a 40-fold change in its orbital angular momentum that is linked with minima and maxima in the 11-year sunspot cycle [21].

This cycle "*is actually half of a 22-year magnetic cycle (Hale cycle), during which the polarity of sunspot groups reverse twice, hence returning to its original state*" [ref. 21, p. 415]. Sunspots regularly appear at moderate latitudes and migrate toward the equator [22]. Measurements of the solar wind and solar magnetic fields reveal that "*On average, solar magnetic field lines in the ecliptic plane point outward on one side of the Sun and inward on the other, reversing direction approximately every 11 years*" and "*The data are consistent with a model in which the solar magnetic dipole returns to the same longitude after each reversal*" [ref. 23, p. 2315].

Magnetic fields drive solar activity, and the solar wind carries the Sun's magnetic field to Earth. However a 38-year compilation of measurements by Earth-orbiting satellites and interplanetary spacecraft, including about 335,000 hours of solar wind speed data and 250,000 hours of magnetic field data, failed to explain the origin of these magnetic fields in terms of the SSM [23]. Some authors acknowledge that an internal magnetic field may exist within the Sun, e.g.,

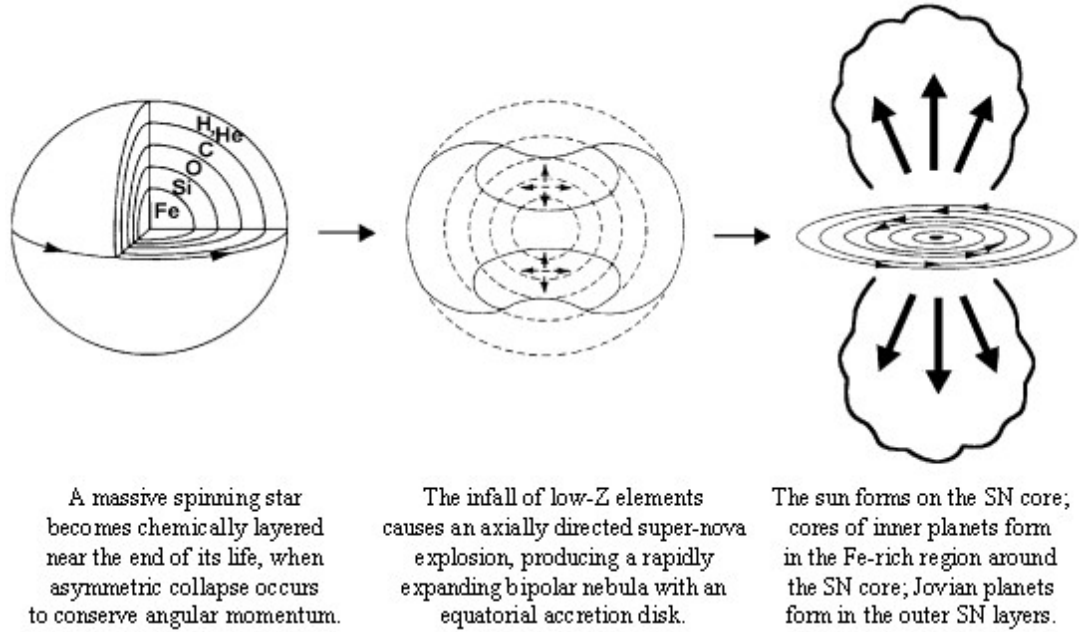

A massive spinning star becomes chemically layered near the end of its life, when asymmetric collapse occurs to conserve angular momentum.

The infall of low-Z elements causes an axially directed super-nova explosion, producing a rapidly expanding bipolar nebula with an equatorial accretion disk.

The sun forms on the SN core; cores of inner planets form in the Fe-rich region around the SN core; Jovian planets form in the outer SN layers.

**Fig. 1.** Formation of the solar system from debris of a single supernova was proposed in 1975 [27] to explain the decay products of short-lived nuclides and inter-linked chemical and isotopic heterogeneities in meteorites and planets. This figure from ref. [12] offers two viable explanations for the solar magnetic fields that are linked with solar activity [19], with the solar wind, with solar flares, and with other parameters that influence the Earth's climate [20].

[22, 24], and the first author of [23] notes in a recent JPL news report that, "*There may be something asymmetric about the Sun's interior, perhaps a deep-seated lump of old magnetic field*" [25]. Such uncertainties undermine efforts to understand the Sun's effect on the Earth's climate.

The SSM assumes the Sun formed suddenly as a fully convective, homogeneous, hydrogen-rich body, with no accretion of mass [26]. In view of the enormous influence of magnetic fields on solar activity and on the planet Earth, and the apparent inability of the SSM to explain these, it seems auspicious to consider the magnetic fields that might be generated if the Sun instead formed in a timely fashion by accretion of the same elements that formed nearby planets [27]. This scenario is depicted in Fig. 1. Several papers cited earlier [1, 7-15] give other experimental observations which indicate that the Sun is iron-rich and formed in this manner. Manuel and Friberg [1] present the most complete and concise summary of observations unexplained by the SSM [26] that "*have been accumulating like water behind an earthen dam since the early 1960s*" [ref. 31, p. 479].

## III.    THE ORIGIN OF SOLAR MAGNETIC FIELDS

The prevailing opinion in the solar physics community is that solar dynamos generate the Sun's magnetic fields by plasma flows in the outer part of the Sun, the convection zone. The model of a hydrogen-filled Sun offers few other options. The solar tachocline separates the outer convective zone from the inner radiative zone, about $5 \times 10^5$ km from the center of the Sun at a depth of 0.7 $R_o$, where $R_o$ is the solar radius ($7 \times 10^5$ km). Helioseismology shows that the differential flows that characterize the outer 30% of the Sun disappear at the tachocline, and the Sun rotates as a rigid body below about 0.7 $R_o$ [22]. Thus, Ruzmaikin and Lindsey [32] state that, "*The magnetic field is generated by motions (flows) of the conducting plasma in the convection zone*" and "*... the solar dynamo operating in the convective zone generates maximal magnetic fields near the base of the convection zone.*" [ref. 32, p. 71]. Hathaway [33] also notes that "*Large scale flows within the solar convection zone are thought to be the primary drivers of the Sun's magnetic activity cycle*" but acknowledges that "*Understanding the nature of the solar* *activity cycle has been a major problem for solar astronomy in spite of nearly 400 years of observations.*" [ref. 33, p. 87].

A few quotes from Toomre's overview paper [34] at the latest SOHO/GONG Conference illustrate the importance of solar magnetic fields and the need for a better understanding of their origin: "*The high speed solar wind and its energetic particles, the sectoral structure of the helio-sphere, coronal holes and mass ejections, and flares are all linked to the variability of the magnetic fields that pervade the solar atmosphere.*" [ref. 34, p. 3]. Efforts to understand these are "*... made even more difficult by only having access to very simplified models describing the origin of these magnetic fields.*" [ref. 34, p. 3]. "*The global or large-scale dynamo, seated within the strong rotational shear of the tachocline at the base of the convection zone, is responsible for global properties of the magnetic field including cyclic behavior, active regions, and sunspot groups.*" [ref. 34, p. 4]. However new, unanticipated findings "*... are forcing reconsideration of how magnetic fields are generated in the solar interior.*" [ref. 34, p.3].

Formation of the solar system in the manner depicted in Fig. 1 is based on a multitude of other experimental observations that are unexplained by the SSM [1, 7-15]. Zinner [28] presents a recent summary of evidence for the decay products of $^{244}$Pu [29], $^{60}$Fe [30] and other extinct nuclides produced in the supernova explosion about 5 Gy ago [29]. Highly refractory products of early condensation (diamond, graphite, silicon carbide, etc.) trapped high levels of extinct radioactivies and very anomalous isotope ratios from the fresh supernova debris, e.g., [8]. Less refractory condensates were partially destroyed about 4.5-4.7 Gy ago when H-fusion was re-ignited in the Sun and some planetary dust was converted by flash heating into the aerodynam-ically shaped droplets in meteorites called "chondrules" [7-15].

Unlike the SSM, forming an iron-rich Sun in the manner shown in Fig. 1 from highly evolved nuclear material near the core of a supernova offers two plausible explanations for deep-seated, solar magnetic fields from the birth of the Sun:

1. Oppenheimer and Volkoff [6] predict that the mass of a neutron star, such as the one on which the Sun formed (Fig. 1), will constitute 33-75% of the total mass of the Sun. Rotating neutron stars (pulsars) are $\approx 10$ km objects with typical surface magnetic fields around $10^{12}$ G [35].

2. Ninham [2] predicts Bose-Einstein condensation of the dense plasma into a superconductor for the highly evolved nuclear material produced by Type I and II supernovae (Fig. 1). The magnetic field is expelled to the surface of superconductors by the Meissner effect, and Ninham notes that "*If a transition to a condensed state occurs, then one can be certain that magnetic effects associated with the transition can not be neglected, and will be most unusual.*" [ref. 2, p. 279]. This rotating superconducting, superfluid will generate a magnetic field continuously and expel it to the surface of the outer core.

Depending on their lifetimes, either or both of these sources of magnetic fields in the interior of the Sun may be linked with solar activity [19], including the overall solar climate, the solar wind, solar flares, and many other parameters that affect Earth and its climate [20].

Gold notes that the magnetic field from a pulsar *"will be down to values the order of $10^3$-$10^4$ gauss"* at a distance of 75-400 x $10^3$ km [ref. 35, p. 732]. Corbard [24] states that a magnetic flux density of $10^3$ - $10^4$ G would confine the tachocline to its observed thickness, < 0.05 $R_o$. The azimuthal component of the field is estimated to be 3 x $10^5$ G, or about 100 times larger than the radial component near the solar tachocline [32]. The solar tachocline is about 500 x $10^3$ km from the center of the Sun. Thus, a central neutron star alone could account for a significant fraction of the observed solar magnetic fields.

The other magnetic field, expelled by the Meissner effect to the surface of the superfluid, superconductor made by Bose-Einstein condensation of iron-rich, zero-spin nuclides [2], would be closer to solar surface and would likely have an even greater affect on the solar cycle and events at the solar surface that impact the climate of nearby planets.

## IV.    CONCLUSIONS, TESTS, AND RELATED OBSERVATIONS

It may be helpful to refer to the formation of the Sun and its planetary system depicted in Fig. 1 if questions arise concerning any of the following conclusions, tests, or related observations.

Conclusions:

- Ancient magnetic fields in the deep interior of the Sun likely arise from a) the collapsed supernova remnant [35] at the solar core [12,13], and/or b) the iron-rich, super-conducting material [2] that surrounds the neutron star [7-11]. Fig. 1 shows how these were produced.

- The long-term memory of solar magnetic fields [21-25] is most likely caused by these deep-seated fields established at the birth of the solar system (Fig. 1).

- These magnetic fields may cause many of the intermittent changes that disrupt our climate.

- Deep-seated, solar magnetic fields accelerate $H^+$ ions upward, maintaining mass separation at the solar surface and in the quiet solar wind. $H^+$ ions not consumed by fusion in this upward journey then depart in the solar wind at an annual rate of 3 x $10^{43}$ per year [1, 11-15 ].

- Magnetic fields that penetrate the solar surface produce many other global properties of the Sun: The 22-year solar magnetic cycle, coronal holes, mass ejections, and the more energetic, less mass-fractionated material observed in solar flares [1, 11-15 ].

- Magnetic fields likely transport angular momentum upward from the source, reaching the surface at mid latitudes as sunspots that migrate towards the equator. This is probably why equatorial material rotates $\approx$35 % times faster than polar material at the solar surface [22].

- Superfluidity and quantized vortices in nucleon-paired Fermions of the solar core may also cause abrupt changes in the Sun, like those seen in pulsars [3], and perhaps even induce "glitches" in the Earth's magnetic field [17,18].

- The central neutron star likely contributed to luminosity in the early Sun and may still do so today. Gold [36] notes that "*The rotational energy of a newly formed neutron star is a major fraction of the energy released in its collapse, and it is of the same order as the total energy the star ever had available from nuclear sources.*" [ref. 36, p. 27].

- The gravitational field from a central, massive neutron star may also explain how the Sun maintains an overall spherical shape, with less ellipticity (oblateness) than the Earth, despite fluidity that permits differential rotation in the outer 30% by radius [37].

Proposed Tests

- If deep seated magnetic fields selectively transport $H^+$ ions (fermions) upward, those that penetrate the surface might be enriched in other fermions and could be identified as isotopic excesses of other, odd-A nuclides like $^3$He, $^{21}$Ne, $^{25}$Mg, $^{83}$Kr, $^{131}$Xe, etc., in solar flares.

- Measurements of solar anti-neutrinos with E < 0.782 MeV could be used to test if neutron decay in the Sun is the source of the hydrogen at its surface. Nuclides like $^3$He, $^{14}$N, or $^{35}$Cl might be used as targets to look for inverse beta-decay induced by solar anti-neutrinos. For example, a search for 87-day $^{35}$S from the $^{35}$Cl —> $^{35}$S reaction in the Homestake Mine [38] might confirm or deny the occurrence of neutron decay inside the Sun.

Related Observations

- The Sun is much more heterogeneous [1] and dynamically active [19] than expected from the standard solar model [26].

- Its magnetic activity varies over the solar cycle [21] and violently ejects material in flares with more heavy elements and more heavy isotopes than in the quiet solar wind [1, 11-15].

- The Galileo probe into Jupiter demonstrated that deuterium burning in the Sun [39] cannot explain the presence of mono-isotopic $^1$H and the high value of the $^3$He/$^4$He ratio in the solar wind [40]. These observations provide additional evidence that a) neutron-decay is the likely source of hydrogen at the solar surface and b) mass separation in the Sun enriches lighter nuclides at the solar surface.

Over 25 years ago leading scientific journals published evidence that the Sun is iron-rich [7] and oscillates like a pulsar [41]. The present paper confirms that these two feature of the solar interior may help solar physicists and climatologists elucidate solar magnetic fields and their effect on the solar cycle and the terrestrial climate.

One caveat: We have not shown conclusively that the Sun's inner core is a neutron star, although this offers a plausible explanation for solar luminosity and for the continuous outflow of neutrinos and mono-isotopic $^1$H$^+$ ions from the solar surface [12, 13]. The mechanism by which a magnetic field is generated in pulsars [4] involves a postulated mechanism of pairing of neutrons (fermions) by analogy with electron pairing in metals at very low temperatures. This may well be so, but in metals this (electron) pairing comes about because of electron-phonon coupling, with the required phonons and lattice vibrations being due to particular properties of the underlying metal ion lattice in special low temperature ranges. Whether or not this analogy is valid for astrophysical systems and can so account for magnetic fields through consequent quantized vortices [3] in the Sun, there is little doubt that collapse of the rotating supernova shown in Fig. 1 would generate a magnetic field, concentrated and confined to the surface of the iron-rich Boson fluid, which is well documented [1] and is certainly rotating and condensed [2].

## ACKNOWLEDGEMENTS

This research was supported by the University of Missouri-Rolla (UMR) and the Foundation for Chemical Research, Inc. (FCR). This conclusion to over 40 years of study would not have been possible without the support and encouragement of a few key individuals: Dr. Gary Thomas, the present Chancellor at UMR; the late Mr. Donald L. Castleman, past President of FCR, Inc.; the late Dr. Raymond L. Bisplinghoff, Chancellor at UMR in 1975-1976; the late Professors Paul Kazuo Kuroda, Universities of Tokyo and Arkansas, and John H. Reynolds and Glenn T. Seaborg, University of California-Berkeley.